\documentclass[journal,10pt]{IEEEtran}

\pdfoutput=1

\usepackage{geometry}
\usepackage{graphicx}
\usepackage{hyperref}
\usepackage{mathtools}
\usepackage{bm}
\usepackage[noadjust]{cite}
\usepackage{xspace}
\usepackage{amsmath}
\usepackage{xcolor}
\usepackage{amsfonts}
\usepackage{setspace}
\usepackage{ifthen}
\usepackage{amsmath}

\newboolean{hidenotes}
\newboolean{singlecolumn}

\newcommand{\TODO}[1]{
  \ifthenelse{\boolean{hidenotes}}
  {}{\textcolor{red}{ TODO: #1 }}
}

\newcommand{\comment}[1]{
  \ifthenelse{\boolean{hidenotes}}
  {}{\textcolor{blue}{ #1 }}
}

\newcommand{\revnum}[2]{
  \ifthenelse{\boolean{hidenotes}}
  {}{\marginpar{\textcolor{red}{ R#1.#2 }}}
}

\renewcommand{\vec}[1]{\bm{#1}}

\newcommand{\norm}[1]{\left\lVert#1\right\rVert}
\newcommand{\nth}[1]{#1^{\mathrm{th}}}
\newcommand{\Tone}{$T_1$\xspace}
\newcommand{\Ttwo}{$T_2$\xspace}
\newcommand{\complex}[1]{\mathbb{C}^{#1}}
\newcommand{\real}[1]{\mathbb{R}^{#1}}

\newcommand{\mysection}[1]{\vspace{0pt}\section{#1}\vspace{0pt}}

\DeclareMathOperator{\spn}{span}

\def\sectionautorefname~#1\null{%
  Section~#1\null
}
\def\subsectionautorefname~#1\null{%
  Subsection~#1\null
}
\def\equationautorefname~#1\null{%
  (#1)\null
}

\newcommand{\myfig}[2][placeholder]	%
{
  \begin{figure}[htb]
    \centering
    \ifthenelse{\boolean{singlecolumn}}
    {
      \begin{minipage}[c]{.5\textwidth}
	\includegraphics[width=\textwidth]{figs/#1}
      \end{minipage}\hfill
    }
    {\includegraphics[width=\linewidth]{figs/#1}}
    \ifthenelse{\boolean{singlecolumn}}
    {
      \begin{minipage}[c]{.4\textwidth}
      }
      {\vspace{-8mm}}
      \caption{#2}
    \label{fig:#1}
      \ifthenelse{\boolean{singlecolumn}}
      {
      \end{minipage}
    }
    {}
  \end{figure}
}

\newcommand{\myfigfull}[2][placeholder]	%
{
  \begin{figure*}[htb]
    \centering
    \includegraphics[width=\textwidth]{figs/#1}
    \caption{#2}
    \label{fig:#1}
  \end{figure*}
}

\newenvironment{nalign}{
    \begin{equation}
    \begin{aligned}
}{
    \end{aligned}
    \end{equation}
    \ignorespacesafterend
}

\setboolean{hidenotes}{false} 
\setboolean{singlecolumn}{true}
\setboolean{singlecolumn}{false}
\setboolean{hidenotes}{true}

{\geometry{rmargin=1.25in,lmargin=1.25in,top=1in,bottom=1in}

\begin{document}
\title{Computational MRI with Physics-based Constraints: Application to Multi-contrast and Quantitative Imaging}

\author{
Jonathan I.\ Tamir, Frank Ong, Suma Anand, Ekin Karasan, Ke Wang, Michael Lustig
\thanks{J.I.\ Tamir, S.\ Anand, E.\ Karasan, K.\ Wang, and M.\ Lustig are with the Department of Electrical Engineering and Computer Sciences, University of California, Berkeley. F.\ Ong is with the Department of Electrical Engineering, Stanford University. Corresponding author: J.I.\ Tamir, \url{jtamir@eecs.berkeley.edu}.}
}

\maketitle              %

\ifthenelse{\boolean{singlecolumn}} {\doublespacing}{}
\begin{abstract}
  Compressed sensing  takes advantage of low-dimensional signal structure to reduce sampling requirements far below the Nyquist rate. In magnetic resonance imaging (MRI), this often takes the form of sparsity through wavelet transform, finite differences, and low rank extensions. Though powerful, these image priors are phenomenological in nature and do not account for the mechanism behind the image formation. On the other hand, MRI signal dynamics are governed by physical laws, which can be explicitly modeled and used as priors for reconstruction. \revnum{1}{1}These explicit and implicit signal priors can be synergistically combined in an inverse problem framework to recover sharp, multi-contrast images from highly accelerated scans. Furthermore, the physics-based constraints provide a recipe for recovering quantitative, bio-physical parameters from the data. This article introduces physics-based modeling constraints in MRI and shows how they can be used in conjunction with compressed sensing for image reconstruction and quantitative imaging. We describe model-based quantitative MRI, as well as its linear subspace approximation. We also discuss approaches to selecting user-controllable scan parameters given knowledge of the physical model.
We present several MRI applications that take advantage of this framework for the purpose of multi-contrast imaging and quantitative mapping.


\end{abstract}

\mysection{Introduction}

MRI is a flexible and rich imaging modality with a broad range of applications including visualizing soft tissue contrast, capturing motion, and tracking functional and behavioral dynamics. 
However, long scan times remain a major limitation. A typical MRI
exam consists of several scans, each several minutes long; in comparison, a full-body computed tomography exam takes a few seconds. To lower scan time, concessions are often made in the acquisition process, leading to reduced resolution, image blurring, and lower signal to noise ratio (SNR). Since data collection in MRI consists of Fourier sampling, these tradeoffs can be understood from a signal processing perspective: scan time and SNR increase with the number of Fourier measurements collected, and sampling theory dictates resolution and field of view (FOV).

A recent trend for faster scanning is to subsample the data space while leveraging computational methods for reconstruction. One established technique that has become part of routine clinical practice is parallel imaging \cite{bib:yingspm}, which takes advantage of multiple receive coils placed around the object to acquire data in parallel. Another avenue that is especially suited to MRI is compressed sensing (CS) \cite{bib:lustig2008pe}. In CS MRI, data are acquired below the Nyquist rate using an incoherent measurement scheme such as pseudo-random sampling. A non-linear reconstruction then leverages image structure such as sparsity and low rank to recover the image as if they were sampled at the Nyquist rate.

Though extremely powerful, common image priors in CS fundamentally leverage the natural image statistics,  yet often neglect to leverage the constraints of the imaging physics that create the underlying image in the first place. 
Natural image statistics are phenomenological in nature and do not account for how each voxel (3D pixel) obtained its particular value. Modeling the signal dynamics can help elucidate additional structure in the image, as well as remove unwanted artifacts that are not related to under-sampling. For example, \autoref{fig:fse_blurring_images} shows a high-resolution ``gold-standard" image of a volunteer's foot obtained with a spin-echo sequence over 12 minutes. In comparison, the middle image simulates a 48-second fast spin-echo (FSE) acquisition \cite{bib:hennig1986}, one of the most common sequences used clinically. As the name suggests, FSE is considerably faster, but leads to tissue-dependent image blurring due to signal decay during the acquisition. Even when combined with CS, blurring due to FSE persists, because the sparsity does not address the mechanism behind the blurring.
As we will show in this article, by modeling and exploiting structure in the signal dynamics as well as leveraging CS, it is possible to mitigate the blurring and obtain sharp images \cite{bib:t2shmethod}, shown by the right-most image in \autoref{fig:fse_blurring_images}.
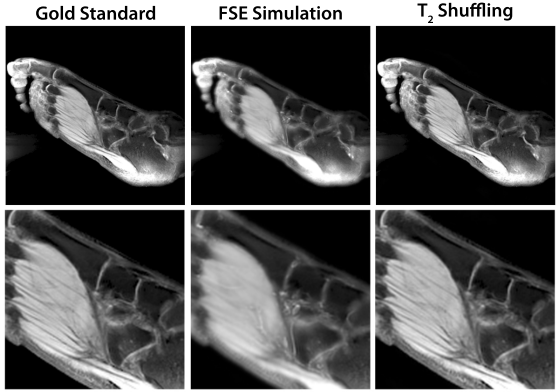
\myfig[fse_blurring_images]{Comparison of (left) a gold-standard spin-echo image acquired in 12 minutes, (middle) an image obtained from a simulated 48-second FSE acquisition, and (right) an image from a simulated 48-second FSE acquisition reconstructed by accounting for and exploiting the a priori signal dynamics that occur during acquisition (\Ttwo Shuffling \cite{bib:t2shmethod}).}

In this article, we provide an overview of computational MRI methods that incorporate physics-based constraints and show how they can be used to accelerate acquisitions, eliminate artifacts, and extract quantitative bio-physical information from the scan.
To simplify the exposition, we limit our scope to spin density and relaxation effects, as these are the most commonly used contrast mechanisms in clinical MRI.\revnum{2}{1} In general, the concepts can be applied to other intrinsic tissue parameters (e.g. diffusion, spectroscopy, chemical exchange, and others), so long as an appropriate physical signal model is included. %
Importantly, many of these approaches have proceeded beyond the ``bench-top'' testing phase, and are actively used in clinical practice.
\revnum{2}{2}The growing interest in characterizing the underlying bio-physical tissue properties follows the larger trend toward personalized, quantitative medicine. Compared to conventional imaging, quantitative MRI can potentially better aid in identifying abnormal tissue, evaluating patients in longitudinal studies, discovering novel biomarkers, and more.\revnum{2}{5}

\section{MRI Physics}\label{sec:physics}

\subsection{Spin Dynamics}\revnum{3}{1}
The MRI system can be approximated by a dynamical system based on the Bloch equations \cite{bib:bloch1946qs}, introduced by Felix Bloch and illustrated conceptually in \autoref{fig:mri_dynamical_system}. An aggregate of spins in each voxel creates a net magnetization that is initially aligned with the scanner's main magnetic field, and evolves based on intrinsic biophysical tissue parameters and user control inputs consisting of radio-frequency (RF) pulses and magnetic field gradients. RF pulses act to rotate the magnetization vector away from the main (longitudinal) field direction and toward the transverse plane. The strength and duration of the RF pulse determine the degree of rotation, or \textit{flip angle}, experienced. The transverse component of the magnetization is sensed by nearby receive coils through Faraday's Law of Induction. We will denote the transverse magnetization as $M_{xy}$ and the longitudinal magnetization as $M_z$.
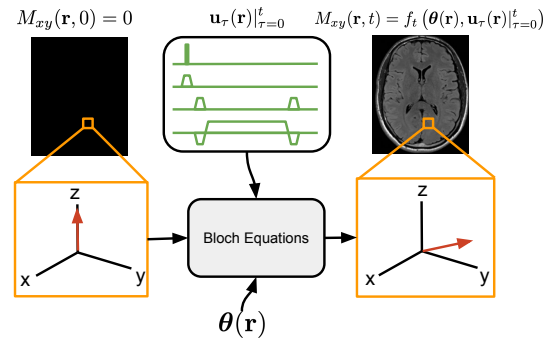
\myfig[mri_dynamical_system]{The MRI dynamical system, governed by the Bloch equations. An aggregate of spins at each voxel position creates a net magnetization, initially pointed in the longitudinal direction, that evolves based on user pulse sequence control inputs and intrinsic tissue parameters. The acquired image is the transverse component of the magnetization.}

Although many tissue parameters influence the signal evolution, here we limit our scope to relaxation parameters, which are the most common contrast mechanism used in MRI.
Relaxation is a fundamental component of nuclear magnetic resonance and dictates the rate that magnetization returns to the equilibrium state, effectively ``resetting'' the MRI system. Longitudinal magnetization exponentially recovers to its initial state with time constant \Tone, and transverse magnetization exponentially decays to zero with time constant \Ttwo.

As \autoref{fig:Bloch_Toilet} shows, relaxation can be understood through an analogy with a toilet\footnote{Introduced by Al Macovski in the 2009 ISMRM Lauterbur Lecture.}, where the water in the tank represents longitudinal magnetization, water in the bowl represents transverse magnetization, and the toilet flush is an RF excitation.
When the toilet is flushed (excitation), water transfers from the tank to the bowl, producing a detectable signal. As the toilet bowl drains (\Ttwo relaxation), the tank refills with water (\Tone recovery). Successive flushes will transfer new water from the tank to the bowl.
Neglecting other system effects, the magnetization evolution after a 90$^\circ$ RF pulse is given by
\begin{align}
  M_z(t) &= 1 - e^{-\frac{t}{T_1}}, \label{eqn:t1} \\
  M_{xy}(t) &=  e^{-\frac{t}{T_2}} \label{eqn:t2},
\end{align}
where $M_{z}(t)$ and $M_{xy}(t)$ are the longitudinal and transverse magnetization components at time $t$ after the excitation, respectively, and have initial conditions of $M_z(0) = 0$ and $M_{xy}(0) = 1$ immediately following the RF pulse.\revnum{3}{2}

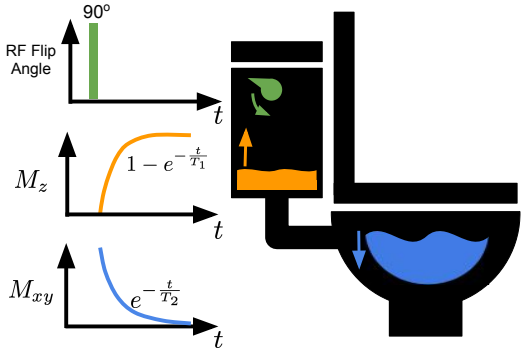
\myfig[Bloch_Toilet]{Signal relaxation visualized through the toilet analogy. The water in the tank represents longitudinal magnetization, water in the bowl represents transverse magnetization, and the toilet flush is an RF excitation. When the toilet is flushed (excitation), water transfers from the tank to the bowl, producing a detectable signal. As the toilet bowl drains (\Ttwo relaxation), the tank refills with water (\Tone recovery). Successive flushes will transfer new water from the tank to the bowl.}

In general, the transverse magnetization distribution in space and time can be described by
\begin{align}
  M_{xy}(\vec r, t) = \rho(\vec r)f_t\left(\vec \theta(\vec r), \vec u_\tau(\vec r)|_{\tau=0}^t\right),
  \label{eqn:mri:mxyt}
\end{align}
where $\vec r \in\real{3}$ represents the 3D spatial position, $\rho(\vec r)$ is the intrinsic amount of magnetization at position $\vec r$ (called proton-density, or PD), and $f_t(\cdot)$ is a spatio-temporal signal evolution that depends on both the biophysical tissue parameters given by $\vec \theta(\vec r)$ and the full history of user-controllable pulse sequence and scanner parameters\footnote{Although the pulse sequence parameters may not have a spatial component, the scanner hardware may introduce a spatial dependence, e.g. due to spatial inhomogeneities.} given by $\vec u_\tau(\vec r)$, $0 \leq \tau \leq t$.\revnum{1}{3}
Although there are numerous tissue parameters and user controls that influence the magnetization, we will focus on the case of FSE imaging, in which the magnetization is sampled at $T$ equispaced intervals with spacing $T_s$. In this case, the image of interest at position $\vec r$ and state $i$ is the sampled transverse magnetization and is given by 
\begin{align}
    x_i(\vec r) = M_{xy}(\vec r, i T_s) = \rho(\vec r)f_{iT_s}\left(\vec \theta(\vec r), \vec u_\tau(\vec r)|_{\tau=0}^{iT_s}\right).
    \label{eqn:fse_signal_evolution}
\end{align}

For FSE, the magnetization is primarily sensitive to relaxation parameters, i.e., $\vec \theta = \left( T_1, T_2 \right)$, and to RF refocusing flip angles, i.e. $\vec u_{iT_s}(\vec r) = \mathrm{RF}_i(\vec r)$, where $\mathrm{RF}_i$ represents the flip angle of the $\nth{i}$ RF pulse.\revnum{1}{4}
Although the RF pulse is not prescribed for each position, in practice the flip angles smoothly vary spatially due to RF transmit field inhomogeneity effects that impart varying levels of RF power across the imaging volume.
We represent the vector of magnetization points at the echo times by $\vec f\left(\vec \theta(\vec r), \vec u(\vec r)\right) \in \complex{T}$, where the $\nth{i}$ component of $\vec f$ is equal to $f_{iT_s}$ (i.e. the transverse magnetization signal).\revnum{1}{5}
Based on the sequence timing and RF flip angle inputs, different types of image contrasts can be created. \autoref{fig:brain_multicontrast} shows four common FSE image contrasts primarily due to PD, \Tone, and \Ttwo, and created by using different sequence parameters in independent scans.
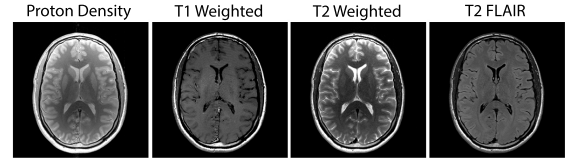
\myfig[brain_multicontrast]{Different image contrasts based on PD, \Tone, and \Ttwo produced by careful choice of RF flip angles and sequence timing for FSE-based scans.}

In particular, $\vec f\left(\vec \theta(\vec r), \vec u(\vec r)\right)$ can be modeled by solutions to the Bloch equations
\cite{bib:bloch1946qs}, which are differential equations that describe the magnetization evolution as a function of time\footnote{In fact the Bloch equations themselves are also phenomenological!}. The Bloch equations allow us to calculate the magnetization (signal) evolution for individual spins in a spatial region, given pulse sequence inputs. For example, the relaxation behavior given by \autoref{eqn:t1} and \autoref{eqn:t2} is the solution to the Bloch equations when no time-varying fields are present. Many extensions have been introduced to model additional contrast mechanisms, including diffusion %
and chemical exchange, %
though these are beyond our scope.

\subsection{Simulating Spins}
Simulating the magnetization evolutions of spins can be a powerful tool for developing and evaluating physics constrained reconstruction methods. For example, in Section~\ref{sec:relax_mri}, we will show how to use simulated signal evolutions to construct an approximate linear subspace for reconstruction. The Bloch equations are linear differential equations, and their numerical simulation can be efficiently computed under reasonable assumptions and approximations.\revnum{3}{4}
In particular, when the RF pulses are discretized, spin simulation using the Bloch equations reduces to successive applications of rotation followed by relaxation to the magnetization vector $\vec{M} = (M_{xy}, M_z)$. Figure~\ref{fig:simulation} shows an illustration of the simulation process. The overall signal evolution is non-linear but differentiable with respect to the relaxation and system parameters.\revnum{2}{3}

\myfigfull[simulation]{When the RF pulses are discretized, spin simulation using the Bloch equations reduces to successive applications of rotation followed by relaxation to the magnetization vector $\vec{M}$. An alternative to the Bloch equations is the extended phase graph (EPG) formalism, which simulates the signal evolution of a distribution of spins across a voxel. EPG efficiently keeps track of signal evolutions across multiple resonant frequencies using the Fourier series.}

In real imaging applications, multiple resonant frequencies often appear within each voxel due to local magnetic field inhomogeneities.\revnum{1}{6} That is, the magnetic field experienced by spins within a voxel can be slightly different. For a realistic simulation, the Bloch equations have to be solved for each resonant frequency separately that appears within one particular voxel, and then summed together to form the resulting signal. This can be computationally demanding.

An alternative to the Bloch equations is the extended phase graph (EPG) formalism, which simulates the signal evolution on a voxel level~\cite{bib:weigel2015}. EPG uniformly discretizes the resonant frequencies within one voxel. The discretization allows EPG to efficiently keep track of signal evolutions across multiple resonant frequencies using the Fourier series. Similar to the Bloch equations, spin simulation using the EPG involves successive applications of rotation followed by relaxation to the underlying magnetization representation. The overall signal evolution using EPG is also non-linear but differentiable.

\section{MRI Sampling}
\subsection{Spatial Encoding}
The received MRI signal represents spatial frequencies of the transverse magnetization distribution in space, and the linear relationship is described by the integral 
\begin{align}
  s(t) = \int_{\vec r} M_{xy}(\vec r, t) e^{-j 2 \pi \vec{k}(t)^\top\vec r} d\vec{r} + w(t),
  \label{eqn:signal_equation}
\end{align}
where $s(t) \in \mathbb{C}$ is the acquired signal at time $t$, $\vec{k}(t) \in\real{3}$ is a trajectory through the 3D frequency space, and $w(t)$ is complex-valued white Gaussian noise. The symbol $(\cdot)^\top$ denotes the transpose operation and $j=\sqrt{-1}.$
Because the spatial frequency wave-number is typically denoted as $\vec k$, MRI acquisitions are often described as sampling in \textit{k-space} \cite{bib:fessler_model_based}.

Throughout this article we consider a discrete Fourier approximation of \autoref{eqn:signal_equation}. 
Though clinical MRI systems come standard with parallel imaging receive arrays \cite{bib:yingspm}, we will omit their discussion for brevity, with the understanding that they can be flexibly incorporated into the sampling model.
We first consider the acquisition model when the magnetization stays constant over time, i.e. $M_{xy}(\vec r, t) = M_{xy}(\vec r)$ with no relaxation effects. Then, given an underlying image $\vec x \in \complex{N}$, which consists of the transverse magnetization of all $N$ voxels, the full forward model is represented in matrix form as
\begin{align}
    \vec y = \vec P \vec F \vec x  + \vec w,
    \label{eqn:forward_model_multicoil_full}
\end{align}
where $\vec y\in\complex{M}$ are the acquired k-space measurements, $\vec F\in\complex{N\times N}$ is a discrete Fourier transform operator, $\vec P \in \complex{M\times N}$ is a sampling operator that selects the acquired k-space measurements, and $\vec w \in \complex{M}$ is the noise.
The encoding operator is succintly represented as $\vec E = \vec P \vec F$.

Since the scan time is directly proportional to the number of measurements, we are typically interested in solving problems for the case where $M < N$. Compressed sensing offers an avenue for targeting this regime by exploiting low-dimensional structure in the image representation \cite{bib:lustig2008pe}. A common inverse problem approach to CS MRI is a regularized least-squares optimization given by
\begin{align}
\underset{\vec x}{\arg\min}\frac{1}{2}\norm{\vec y - \vec E \vec x}_2^2 + \lambda \mathcal{R}(\vec x),
  \label{eqn:lasso}
\end{align}
where $\mathcal{R}$ is a sparsity-promoting regularization, e.g.\ $\ell_1$ norm of  the wavelet coefficients, and $\lambda>0$ is a regularization term.

\subsection{Multi-dimensional Extensions}
Although initial work on CS MRI focused on sparsity of static, anatomical images using spatial wavelet transforms and total variation \cite{bib:lustig2008pe}, many extensions have been proposed to handle additional imaging dimensions, including joint sparsity, low rank, and their variants \cite{bib:trzasko2013dw, bib:frankmlr, bib:christodoulou2018}.
In particular, the linear forward model can be extended to represent additional image states, $\vec x = \begin{bmatrix}\vec x_1 & \cdots & \vec x_T\end{bmatrix}$, where $\vec x_i \in \complex{N}$ is the image at the $\nth{i}$ image state and $T$ is the number of states. The forward model, illustrated in \autoref{fig:forward_model_SA} for signal relaxation, includes different encoding operators for each state $i$ 
based on the user-specified sampling patterns:
\begin{align}
  \vec E_{i} = \vec P_i \vec F.
  \label{eqn:temporal_forward_model}
\end{align}
This  concept can be used to include additional dimensions representing signal relaxation \cite{bib:t2shmethod}, cardiac and respiratory motion \cite{bib:lustig2008pe,bib:feng2016zk,bib:jiang2018pe}, and many others \cite{bib:christodoulou2018}. Selecting the $T$ sampling patterns in a way that maintains compatibility with parallel imaging and CS is an active research area \cite{bib:levine2018jk, bib:haldar2018nj}.
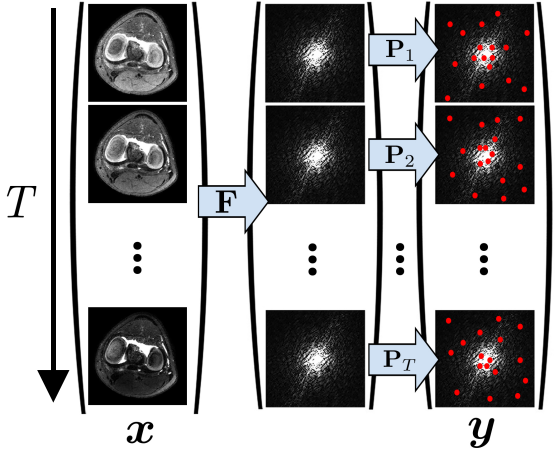
  \myfig[forward_model_SA]{The forward model extended to temporal relaxation. Each image represents a sampling time $i$ during the acquisition with $T$ time points, where each time point is Fourier transformed and sampled with a different sampling operator $\vec P_i$ (represented by red circles).}

\section{Model-based MRI}\label{sec:model_based_mri}
The signal evolution model developed in the previous sections describes how the signal received by the MRI scanner is formed. Combining with spatial encoding, the physics based forward model considers the following non-linear evolution:
\begin{nalign}
    x_i(\vec r) &= \rho(\vec r)f_{iT_s}\left(\vec \theta(\vec r), \vec u(\vec r)\right),\\
    \vec{y}_i &= \vec{E}_i \vec{x}_i, \quad i=1,\dots,T.
    \label{eqn:physics_based_forward_model}
\end{nalign}
The most explicit use of the physics model in a reconstruction is to directly solve for the tissue parameters from the raw k-space measurements. This is in contrast to first reconstructing a time series of images followed by a parameter fit. It is 
possible to formulate this model-based inversion even in the case of undersampled k-space  \cite{bib:block2009,bib:zhao2014b}. 
This can be written as a non-linear, non-convex least squares objective in which we aim to solve for ($\vec \rho$, $\bm \theta$): 
\begin{nalign}
    & \underset{\vec \rho, \bm \theta}{\text{minimize}}
& & \frac{1}{2} \norm{\vec{E} \vec{x} - \vec{y}}_2^2 + \lambda \mathcal{R}(\vec \rho, \bm \theta)\\
& \text{subject to}
& & x_i(\vec r) = \rho(\vec r)f_{iT_s}\left(\vec \theta(\vec r), \vec u(\vec r)\right),\\
& & & i=1,\dots,T.
\label{eqn:model_based_recon}
\end{nalign}
In the MRI literature, equation~\autoref{eqn:model_based_recon} is often referred to as quantitative MRI with a model-based reconstruction, as the physical model is incorporated into the objective function.

Compared to the conventional image series reconstruction followed by a fit, solving for the parameters directly serves as the ultimate dimensionality reduction: instead of computing a set of images, the aim is to recover only the intrinsic information represented by $\vec \rho$ and $\vec \theta$. The main benefit of this approach is that the problem size is significantly reduced. For example, for mapping PD and \Ttwo, solving a problem of the form \autoref{eqn:lasso} consists of $TN$ unknown variables, while the model-based representation \autoref{eqn:model_based_recon} has only $2N$ unknowns. In addition, sparsity-promoting regularization penalties can be applied directly to the parameter maps. 
Since the data fidelity term is differentiable, the overall problem can be optimized using first-order or second-order methods.

Additional system-related parameters such as RF field inhomogeneity and off-resonance can also be modeled and used as variables to be solved for as part of the reconstruction. Explicit incorporation of these systematic deviations can make the reconstruction more robust to imperfections. With the parameters in hand, synthetic contrast-weighted images could be potentially generated by evaluating \autoref{eqn:mri:mxyt} with specific scan and sequence parameters, a technique known as Synthetic MR
\cite{bib:block2009,bib:tanenbaum2017kd}.

However, even though the individual terms are differentiable, the main downside is that the reconstruction problem is highly non-convex, thereby complicating the optimization.  The non-convexity results in increased computation time and dependence on initialization.
The problem requires a good initial guess of $\vec{\rho}$ and $\vec \theta$ in order to converge to a reasonable estimate. Another drawback is that mismatches between the model and the true acquisition can lead to error propagation in the estimated parameter maps. \revnum{2}{4}For example, radiologists have described the presence of flow artifacts, white-noise artifacts, and other artifacts in synthetic FLAIR contrast images \cite{bib:tanenbaum2017kd}, likely due to unmodeled effects such as flow. Partial voluming is also known to impair the estimation accuracy \cite{bib:block2009}. Errors in sampling trajectories due to eddy currents and gradient delays can also manifest as blurring and streaking artifacts.
These effects can potentially be reduced by expanding the signal model, e.g. to incorporate scanner non-idealities. The pulse sequence can also include navigator components to aid in the estimation.

\section{Relaxing the Multi-contrast Model}\label{sec:relax_mri}
Instead of fully incorporating MRI physics into the reconstruction, it is possible to use  relaxed constraints that are more amenable to optimization \cite{bib:doneva2010xi}. The natural dynamics of the MR signal are constrained by the Bloch equations, implying the existence of a low-dimensional manifold, which is non-linear in general.
As \autoref{fig:manifold_subspace_redo} illustrates, a linear subspace approximation to the manifold with higher dimensionality could provide a compromise between representation simplicity and size.
This may be attractive for a few reasons; namely, to maintain convexity and computational efficiency in the optimization, to decouple the reconstruction from the quantitative fitting, and to reduce propagation of model error. In addition, many applications do not require quantitative parameters, and instead rely on high-quality contrast-weighted images. This is the case for nearly all clinical diagnostic imaging. Unfortunately, contrast-weighted images derived from parameter maps are susceptible to error propagation due to unmodeled components, e.g. from partial voluming and flow effects, and have seen limited use clinically \cite{bib:tanenbaum2017kd}.
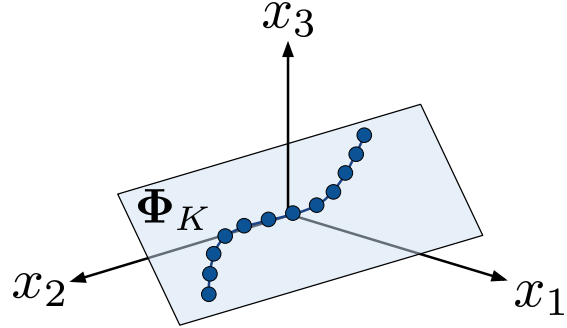
\myfig[manifold_subspace_redo]{The low-dimensional manifold representing the Bloch equations is captured by a linear subspace $\vec \Phi_K$ of larger dimension.}

\subsection{Subspace Constraint}
Continuing with the FSE sequence as a guiding example, \autoref{fig:fse_sim_white} shows the signal evolutions for an FSE simulation with a particular flip angle schedule. Despite differences in relaxation parameters, the signal evolutions for different tissues follow similar trends. This correlation implies low-dimensional structure; namely, the signal evolutions of different tissues form a low-dimensional subspace \cite{bib:liang2007ye,bib:huang2012oe, bib:zhao2014, bib:t2shmethod,bib:velikina2015jx}.
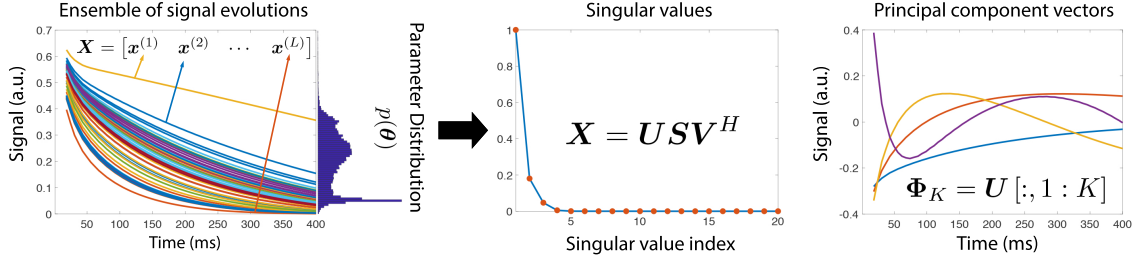
\myfigfull[fse_sim_white]{Forming a subspace based on signal dynamics, shown for FSE. (left) An ensemble of signal evolutions is drawn from a prior distribution. (middle) Due to correlation in signal dynamics, the data matrix is low-rank and (right) is well-approximated with PCA.}

Many approaches can be taken to design the subspace. Here we focus on simulating a set of training signals derived from EPG simulation. We assume a prior distribution $p(\vec \theta)$ is known and draw $L$ samples from the distribution to create the training signals. The prior distribution can be taken from known literature values, or from a conventional mapping procedure focused on a particular anatomy.

Consider a data matrix $\vec X \in\complex{T\times L}$ consisting of an ensemble of $L$ signal evolutions sampled at $T$ echo times. Each column in $\vec X$ represents
the signal evolution of a spin population with a particular $\vec \theta \in p(\bm \theta)$. Let $\vec \Phi\in \complex{T\times T}$ be an orthonormal
temporal basis, i.e.
\begin{align}
  \vec X = \vec\Phi\vec\Phi^H\vec X.
  \label{eqn:t2sh:span}
\end{align}
The goal is to design $\vec\Phi = \begin{bmatrix}\bm\varphi_1 & \cdots & \bm\varphi_T\end{bmatrix}$ and a $K-$dimensional subspace, \sloppy
$\vec\Phi_K = \spn\{\vec\varphi_1, \dots, \bm\varphi_K\}$, such that
\begin{align} \norm{\vec X - \vec \Phi_K\vec\Phi_K^H \vec X} < \epsilon,
  \label{eqn:t2sh:span2}
\end{align}
where $\epsilon$ is a modeling error tolerance.
The choice of norm in \autoref{eqn:t2sh:span2} will affect the chosen subspace and can be used to capture average, worst-case, and other error metrics.

When the Frobenius norm is used, the solution corresponds to the truncated singular value decomposition
\cite{bib:liang2007ye,bib:huang2012oe,bib:t2shmethod}, i.e. principal component analysis (PCA), where $\vec \Phi_K$ consists of the left singular vectors of $\vec X$ corresponding to the $K$ largest singular values. The principal component images, corresponding to the principal component vectors, are given by
\begin{align}
    \vec \alpha = \vec\Phi_K^H \vec x.
    \label{eqn:pca}
\end{align}
Based on \autoref{eqn:t2sh:span2}, we also have $\vec x \approx \vec \Phi_K \vec\alpha$.

\subsection{Reconstruction}
The subspace relationship can be incorporated into the reconstruction as additional prior knowledge in the form of regularization \cite{bib:velikina2015jx}:
\begin{align}
  \min_{\vec x} \frac{1}{2}\norm{\vec y - \vec E\vec x}_2^2 + 
  \frac{\mu}{2}\norm{\vec x - \vec\Phi_K\vec\Phi_K^H \vec x}_2^2 
  + \lambda \mathcal{R}\left(\vec x\right),
  \label{eqn:mocco}
\end{align} 
where $\mu>0$ controls the degree of subspace modeling. Alternatively, the subspace can be used as an explicit constraint, i.e.,
\begin{nalign}
    & \underset{\vec x}{\text{minimize}}
& & \frac{1}{2} \norm{\vec y - \vec{E} \vec{x}}_2^2
+ \lambda \mathcal{R}\left(\vec x\right)\\
& \text{subject to}
& & \vec x = \vec \Phi_K \vec \alpha.
\label{eqn:subspace_constrained_recon}
\end{nalign}
Although the analysis form described by \autoref{eqn:mocco} is a more faithful representation, the reconstruction still requires solving for $TN$ parameters in $\vec x$, in addition to introducing a hyper-parameter $\mu$. In contrast, the synthesis form used in \autoref{eqn:subspace_constrained_recon} introduces explicit model error but is significantly more computationally efficient.

When using the hard constraint, we can now solve for the subspace coefficient images directly \cite{bib:t2shmethod},
\begin{align}
  \min_{\bm\alpha} \frac{1}{2}\norm{\vec y - \vec E\vec\Phi_K\bm\alpha}_2^2 + \lambda\mathcal{R}\left(\vec \alpha\right),
  \label{eqn:t2shrecon}
\end{align}
followed by back-projection: $\hat{\vec x} = \vec\Phi_K\vec\alpha$. This is a significant dimensionality reduction! Instead of solving for $TN$ variables, we only need to solve for $KN$ variables. In addition, when the normal equations are used in an iterative optimization (as is the case for many first-order iterative algorithms), we can take advantage of the commutativity of the subspace operator and the Fourier transform, as the former only operates on the parametric dimension and the latter only operates on the spatial coordinates \cite{bib:t2shmethod}:
\begin{align}
  \vec P \vec F \vec \Phi_K & = \vec P \vec \Phi_K \vec F\\
  \implies  \vec \Phi_K^H \vec F^H \vec P \vec F \vec \Phi_K &= \vec F^H\vec \Phi_K^H\vec P\vec \Phi_K\vec F \\
  &= \vec F^H \vec \Psi_K \vec F,
  \label{eqn:stkern}
\end{align}
where $\vec\Psi_K \in \complex{KM\times KM}$ is a block-wise diagonal operator with a $K\times K$ symmetric block  for each spatial frequency point. In other words, the optimization does not have to perform any computation in the ambient space, and the complexity grows with $K$, independent of $T$.

Compared to model-based quantitative mapping, the subspace-constrained forward model is convex and easier to solve. In addition, the subspace is less sensitive to model error, as it does not strictly impose a specific physical model. As an example, a voxel with partial voluming will contain a linear combination of signal evolutions, $\vec x(\vec r) = a_1\vec x^{(1)}(\vec r) + a_2\vec x^{(2)}(\vec r)$, comprising different tissue parameters $\vec \theta^{(1)}$ and $\vec \theta^{(2)}$, which is inconsistent with \autoref{eqn:mri:mxyt}. In contrast, if $\vec x^{(1)}$ and $\vec x^{(2)}$ are separately represented by the subspace, then so is their combination.

The main drawback to the subspace formulation is that the problem is not reduced to its intrinsic dimension governed by $\vec \rho$ and $\vec \theta$. However, in practice a subspace size of $K<5$ is practical for applications even when the ambient dimension is in the hundreds \cite{bib:t2shmethod,bib:zhao2017}.
A second drawback is that the subspace can represent points off the manifold, that are not physically meaningful. Thus, data inconsistencies can manifest as inaccurate images after back-projections.

\subsection{Quantitative Mapping}
A straightforward approach to quantitative mapping from reconstructed multi-contrast images is to perform a voxel-wise non-linear least squares fit based on the signal model. Given the reconstructed image $\hat{\vec x}$ at voxel $\vec r$, this amounts to solving
\begin{align}
  \underset{\rho(\vec r),\bm\theta(\vec r)}{\arg\min}\frac{1}{2}\norm{\hat{\vec x}(\vec r) - \rho(\vec r)\vec f(\bm \theta(\vec r), \vec u(\vec r))}_2^2.
  \label{eqn:nnls}
\end{align}

As with \autoref{sec:model_based_mri}, it is also possible to solve for RF field inhomogeneity by additionally solving for the non-negative scalar $\eta$ that multiplies the flip angles: $\vec u \rightarrow \eta\vec u$.
This formulation also covers dictionary-based methods which have grown in recent popularity with the advent of MR Fingerprinting \cite{bib:ma2013vg}. Rather than solving a continuous non-linear least squares problem, dictionary-based fitting utilizes a grid search across the parameter space and is equivalent to matched filtering. Optimization-based methods can be used as well.%

\revnum{3}{5}When the reconstruction strictly enforces a subspace, there is a known model error between the reconstruction $\hat{\vec x}$ and the signal evolution $\vec{f}$, and we do not expect them to match even in the absence of noise and under-sampling artifacts.
To improve parameter estimation when using a subspace-constrained reconstruction, we can solve for the parameters directly in the subspace:
\begin{align}
\underset{\rho,\vec \theta}{\arg\min}\frac{1}{2}\norm{\hat{\bm \alpha}(\vec r) - \vec \Phi_K^H \rho\vec f(\bm \theta, \vec u(\vec r))}_2^2.
  \label{eqn:nnls_subspace}
\end{align}

Since $\vec f$ is differentiable with respect to $\vec \theta$ and $\vec u$, first-order and second-order solvers can be used. \autoref{fig:t2map} shows an example of solving \autoref{eqn:nnls_subspace} following a \Ttwo Shuffling reconstruction\footnote{Example code available on \url{https://eecs.berkeley.edu/~mlustig/Software.html}} \cite{bib:t2shmethod}, where $K=3$ and  $\bm \theta = (|\rho|, \angle \rho, T_2)$. The non-linear least squares was solved using the Trust Region Reflective algorithm included in the Python Scipy package, and the Jacobians were calculated using the adjoint states method \cite{bib:sbrizzi}.
\revnum{1}{7}
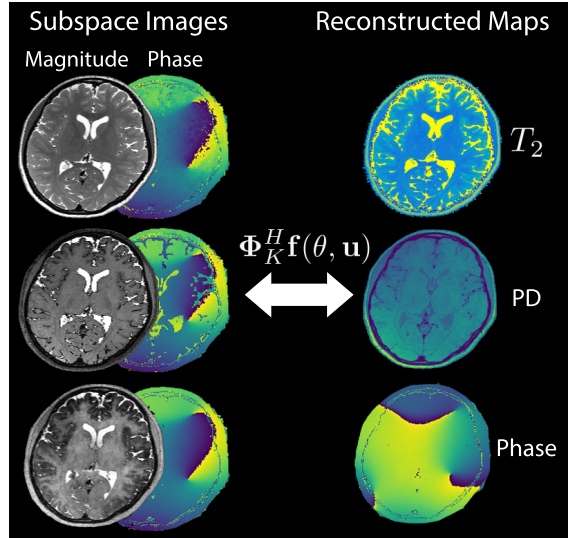
\myfig[t2map]{Example showing voxel-wise parameter fitting following a \Ttwo Shuffling reconstruction using non-linear least squares directly in the subspace. The principal component images (magnitude and phase) were first reconstructed with the convex formulation \autoref{eqn:t2shrecon}, and the complex values were fit to the physical model using the EPG formalism.}

\revnum{2}{6}When solving under-determined inverse problems, it is important to recognize the bias introduced through modeling assumptions and regularization. In particular, the optimization problem \autoref{eqn:subspace_constrained_recon} introduces two forms of bias: model error due to the subspace constraint, and error due to the regularization. The subspace constraint leads to a straightforward tradeoff between model error and noise amplification: noise standard deviation increases with $\sqrt{K}$, where $K$ is the subspace size \cite{bib:t2shmethod}. A small subspace will also reduce sensitivity in the parameter space, and manifests as reduced contrast and parameter-dependent noise amplification, as all voxels are pushed to the same curve fit \cite{bib:huang2012oe}.  
Bias due to regularization can also be an issue in reconstruction and parameter fitting, but its impact depends on the specific regularization used. With non-linear regularization such as $\ell_1$ commonly used in CS, the tradeoff can be harder to quantify. Blocking and smoothing artifacts are common when using wavelet regularization and total variation due to loss of high-frequency content \cite{bib:lustig2008pe}. Low-rank regularization and can lead to washed-out contrast due to loss of spectral information, and its multi-scale variants can lead to blocking artifacts due to their translation-variant structure \cite{bib:frankmlr}. In general, under-regularization can cause residual incoherent under-sampling artifacts and noise amplification, while over-regularization can lead to blurring \cite{bib:block2009,bib:doneva2010xi,bib:huang2012oe,bib:zhao2014,bib:zhao2014b,bib:velikina2015jx,bib:feng2016zk,bib:t2shmethod,bib:zhao2017,bib:jiang2018pe}.

\section{Choosing Scan Parameters}\label{sec:scan_params}
In addition to incorporating MRI physics in the reconstruction, using knowledge of the physics to optimize the MRI pulse sequence could potentially improve the quality of the acquired raw data. Designing an MRI pulse sequence consists of two main components: (\textit{i}) designing the pulse waveform and timing to guide the signal evolution, and (\textit{ii}) designing the spatial encoding to appropriately sample k-space. Initial work focused on designing sequences with simple goals in mind such as maximizing contrast and limiting blurring. A recent trend in computational MRI is to use optimal control to motivate scan parameter selection \cite{bib:sbrizzi}, and incoherent sampling properties to motivate k-space sampling \cite{bib:lustig2008pe}.
Optimal experiment design has also been applied to spatial encoding under a synthesis sparsity model \cite{bib:haldar2018nj}.
Importantly, both spatial encoding and sequence parameters should be designed with a specific reconstruction in mind. %

\subsection{Optimizing Sequence Parameters}
Many approaches to optimizing sequence parameter selection involve the use of a Cramer-Rao lower bound (CRLB), which imposes a lower bound on the variance of an unbiased estimator. To simplify the optimization, the spatial encoding is decoupled from the sequence evolution, and a 1D experiment is used for optimization:
\begin{align}
    \vec y = \vec f\left(\vec \theta, \vec u\right) + \vec w.
\end{align}
Then, the Fisher information matrix, $\vec I(\vec \theta;\bm u)$, which measures the sensitivity of $\vec \theta$ captured by $\vec y$ given sequence parameters $\vec u$,
can be used to lower bound the variance of an unbiased estimator.
The sequence parameters can then be optimized to minimize this lower bound on the variance of the biophysical tissue parameter estimates.
This approach has been used in \cite{bib:Nataraj2016} to develop a general framework to optimize sequence parameters from combinations of pulse sequences for precise estimation of \Tone and \Ttwo parameters jointly. It has also been used to determine sequence parameters for MR Fingerprinting to obtain maximal SNR while not violating physics-based MRI constraints \cite{zhao2016optimal}. 

Another optimal control design based approach uses the EPG formalism to develop a model for the signal evolution in terms of various sequence and tissue parameters \cite{bib:sbrizzi}. An optimization problem is then formulated according to the goal, e.g.\ maximizing signal intensity or minimizing RF power, and solved using the adjoint states method. This approach can flexibly incorporate other objective functions, e.g. to minimize the CRLB, similar to methods discussed above.
\autoref{fig:t2shcrlb} shows an example of optimizing the RF flip angles in an FSE experiment given RF power constraints \cite{bib:keerthivasan2019}. The objective function maximized the component in the Fisher Information matrix corresponding to the variance with respect to \Ttwo estimation. Constant flip angles that achieve the same RF power limit are shown in comparison.
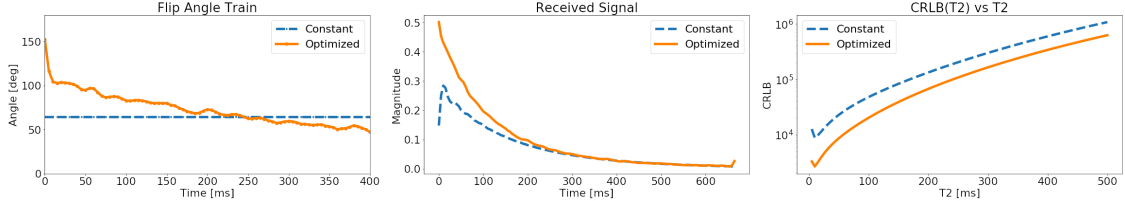
\myfigfull[t2shcrlb]{Comparison of a constant flip angle schedule vs. optimized flip angles for an FSE experiment given a maximum RF power constraint (left).  The CRLB was optimized with respect to $T_1=1000$ ms and $T_2=100$ ms, and the received signal was higher for the optimized flip angles (middle). Even though the optimization targeted a single relaxation value, the bound was lower across a uniform range of \Ttwo values (right).}

Data-driven methods for sequence parameter optimization are also emerging. With these methods, a certain set of sequence parameters are chosen, the resulting time-evolution of the signal for these parameters is either simulated or determined experimentally, and finally an image is reconstructed. The sequence parameters are then updated according to the reconstruction error. Compared to the optimal experiment based approaches, these methods have the advantage of incorporating spatial encoding and reconstruction into the evaluation.

\revnum{3}{6}When choosing sequence parameters, it is important to understand the impact of system imperfections on the optimized sequence. These imperfections can often be incorporated into and mitigated in the optimization framework, though they may greatly increase the computational complexity. For example, to reduce sensitivity to RF transmit field inhomogeneity, one may optimize the average (or min-max) CRLB across spins experiencing a distribution of RF transmit field inhomogeneities by simulating each spin independently. This quickly increases in complexity as additional components are considered.

\subsection{Optimizing Sampling Pattern}
Conventional methods to determine CS sampling trajectories are based on exploiting the incoherence conditions for CS by the use of a transform point spread function (TPSF), which determines the leakage of one transform coefficient to another transform coefficient due to subsampling \cite{bib:lustig2008pe}. 
A sampling pattern can be determined by a Monte-Carlo design procedure where a variable density pattern is obtained by randomly drawing indices using a probability density function, the TPSF of the obtained trajectory is calculated, and the procedure is repeated, choosing the pattern with the lowest peak interference. 
The reconstruction is performed according to \autoref{eqn:lasso} with a sparsity-promoting regularizer on the transform coefficients. Thus, incoherence in the transform domain is a good metric for sampling pattern quality.

Several data-driven approaches have been developed in order to improve on the current incoherence-based methods. These data-driven approaches use a learning-based framework to find the best sampling pattern for a set of training signals. In particular, methods that optimize a sampling pattern for a specific reconstruction rule and anatomy \cite{bib:haldar2018nj} as well as methods that jointly optimize a sampling pattern and a reconstruction strategy have been developed \cite{bahadir2019learning}. In this way, physics-based constraints in the reconstruction are implicitly propagated through to the sampling pattern optimization.\revnum{1}{8}

Although various control design based and data-driven strategies have been developed to optimize one of scan parameters, sampling trajectories and reconstructions, the ultimate goal of developing methods that optimize parameter, trajectory and reconstruction in conjunction remain an area for future work.

\section{Summary}
Since the advent of MRI, physics-based knowledge has been used to recover and understand the signal dynamics and tissue parameters. Many early works have leveraged physical constraints to mitigate image artifacts due to systematic errors and derive quantitative maps. However, long scan time and the lack of sophisticated reconstruction algorithms have prevented the clinical adoption of these techniques.

In the last decade, compressed sensing and other computational imaging approaches have transformed the landscape of what is possible with MRI. Scan time can be appreciably reduced by leveraging natural image statistics in the reconstruction. Many advanced reconstruction algorithms have been developed for this express purpose. Using these numerical tools and combining sparsity-based modeling, it is now feasible to run physics-constrained computational MRI methods in the clinic with reasonable scan and reconstruction times \cite{tamir_targeted_2019}.
By incorporating the physical dynamics due to both tissue-specific and scanner-specific parameters, acquisitions and reconstructions can be designed in tandem to work across a broad patient population in a robust manner.\revnum{2}{2}

The methods introduced in this article provide a framework for modeling the MR dynamics, modifying the acquisition to account for the signal evolutions, and incorporating them into the reconstruction. Although we limited our focus to only modeling dynamics due to relaxation effects and RF pulses, there are many other contrast mechanisms and scanner controls that can be accounted for. On the tissue characterization side, these include diffusion, water-fat imaging, susceptibility, spectroscopy, pharmacokinetics, magnetization transfer, chemical exchange, and blood flow. On the imaging system side, these include field inhomogeneity, eddy current effects, gradient delays, temperature, and more. Each of these components can be modeled jointly, but will greatly increase the dimensionality of the problem. Model-based quantitative imaging reduces the inverse problem to recovering the intrinsic parameters, but careful modeling and simulation must be used to avoid artifacts due to model error. In contrast, subspace constraints and other low-dimensional representations can be used to flexibly capture the dynamics without making strong assumptions.

Many exciting modeling techniques are emerging in the signal processing community, and are great candidates at improving physical modeling in MRI. On the other hand, purely data-driven approaches have grown in popularity, in the hopes of learning the signal characteristics from real data.
\revnum{2}{7}With the growing trend of applying deep learning to inverse problems, there is great promise to incorporating additional physics-based constraints directly in the learning in order to restrict the feasible solution space, reduce the dependence on large training data sets, and model effects not described by the simplified physics \cite{bib:modl}. Like compressed sensing, the rapid empirical progress in deep learning-based imaging should be suitably counter-balanced with theoretical guarantees to guide its use in clinical settings.
Combining physical modeling with data-driven learning is an active area of research in the signal processing community in general, and in the MRI field in particular.

\section*{Acknowledgements}
This work was supported by NIH grant R01EB009690, Sloan Research Fellowship, Bakar Fellowship, and research support from GE Healthcare. The authors thank Volkert Roeloffs, Zhiyong Zhang, and Karthik Gopalan for valuable comments.

\bibliography{refs}

\end{document}